\documentclass{emulateapj}
\usepackage{apjfonts}
\usepackage{xspace}
\usepackage{amsmath}
\usepackage{mathtools}
\usepackage{hyperref} 

\newcommand{\pfrac}{P_{\rm rand}/P_{\rm total} }

\usepackage[usenames,dvipsnames]{xcolor}

\usepackage{etoolbox}
\makeatletter
\patchcmd{\NAT@citex}
  {\@citea\NAT@hyper@{\NAT@nmfmt{\NAT@nm}\NAT@date}}
  {\@citea\NAT@nmfmt{\NAT@nm}\NAT@hyper@{\NAT@date}}
  {}
  {}
\patchcmd{\NAT@citex}
  {\@citea\NAT@hyper@{%
     \NAT@nmfmt{\NAT@nm}%
     \hyper@natlinkbreak{\NAT@aysep\NAT@spacechar}{\@citeb\@extra@b@citeb}%
     \NAT@date}}
  {\@citea\NAT@nmfmt{\NAT@nm}%
   \NAT@aysep\NAT@spacechar%
   \NAT@hyper@{\NAT@date}}
  {}
  {}
\patchcmd{\NAT@citex}
  {\@citea\NAT@hyper@{%
     \NAT@nmfmt{\NAT@nm}%
     \hyper@natlinkbreak{\NAT@spacechar\NAT@@open\if*#1*\else#1\NAT@spacechar\fi}%
       {\@citeb\@extra@b@citeb}%
     \NAT@date}}
  {\@citea\NAT@nmfmt{\NAT@nm}%
   \NAT@spacechar\NAT@@open\if*#1*\else#1\NAT@spacechar\fi%
   \NAT@hyper@{\NAT@date}}
  {}
  {}
\makeatother

\begin{document}

\title{Hydrodynamic Simulation of Non-thermal Pressure Profiles of Galaxy Clusters}
\shorttitle{Non-thermal Pressure Profiles of Galaxy Clusters}
\shortauthors{Nelson, Lau, \& Nagai}
\slugcomment{{\sc Accepted to ApJ:} July 14, 2014}

\author{Kaylea Nelson\altaffilmark{1,3}}
\author{Erwin T. Lau\altaffilmark{2,3}}
\author{Daisuke Nagai\altaffilmark{1,2,3}}

\affil{ 
$^1$Department of Astronomy, Yale University, New Haven, CT 06520, U.S.A.; \href{mailto:kaylea.nelson@yale.edu}{kaylea.nelson@yale.edu} \\
$^2$Department of Physics, Yale University, New Haven, CT 06520, U.S.A.\\
$^3$Yale Center for Astronomy \& Astrophysics, Yale University, New Haven, CT 06520, U.S.A. \\
}

\keywords{cosmology: theory -- galaxies: clusters: general -- methods: numerical }

\begin{abstract}
Cosmological constraints from X-ray and microwave observations of galaxy clusters are subjected to systematic uncertainties. Non-thermal pressure support due to internal gas motions in galaxy clusters is one of the major sources of astrophysical uncertainties. Using a mass-limited sample of galaxy clusters from a high-resolution hydrodynamical cosmological simulation, we characterize the non-thermal pressure fraction profile and study its dependence on redshift, mass, and mass accretion rate. We find that the non-thermal pressure fraction profile is universal across redshift when galaxy cluster radii are defined with respect to the mean matter density of the universe instead of the commonly used critical density.  We also find that the non-thermal pressure is predominantly radial, and the gas velocity anisotropy profile exhibits strong universality when galaxy cluster radii are defined with respect to the mean matter density of the universe.  However, we find that the non-thermal pressure fraction is strongly dependent on the mass accretion rate of the galaxy cluster. We provide fitting formulae for the universal non-thermal pressure fraction and velocity anisotropy profiles of gas in galaxy clusters, which should be useful in modeling astrophysical uncertainties pertinent to using galaxy clusters as cosmological probes.  
\end{abstract}

\section{Introduction}

Clusters of galaxies are the largest gravitationally bound objects in the universe and therefore trace the growth of large scale structure. In recent years, X-ray and microwave observations have enabled detailed studies of the structure and evolution of galaxy clusters and significantly improved the use of these systems as powerful cosmological probes \citep[][for a review]{Allen2011}.  However, current cluster-based cosmological constraints are limited by systematic uncertainties associated with cluster astrophysics. Controlling these astrophysical uncertainties is therefore critical for exploiting the full statistical power of ongoing and upcoming cluster surveys, such as \emph{Planck}\footnote{\url{http://www.rssd.esa.int/index.php?project=planck}} and \emph{eROSITA}\footnote{\url{http://www.mpe.mpg.de/eROSITA}}.

One of the main challenges in using clusters as cosmological probes lies in the accurate determination of their masses. Cluster mass estimates from X-ray and Sunyaev-Zel'dovich (SZ) observations are based on the assumption that cluster gas is in hydrostatic equilibrium with their gravitational potential, but there have been inconsistencies between the hydrostatic mass and the mass estimated from gravitational lensing \citep[e.g.,][]{Zhang2010, Mahdavi2013, vonderLinden2014, Applegate2014}. Hydrodynamical simulations suggest that this hydrostatic mass bias arises from non-thermal pressure support in clusters that is not accounted for in current X-ray and SZ cluster mass measurements \citep[e.g.,][]{Evrard1996, Rasia2006, Nagai2007b, Piffaretti2008}.  Simulations also suggest that accounting for the non-thermal pressure support can recover the cluster mass to within a few percent \citep[e.g.,][]{Rasia2004, Lau2009, Nelson2012, Nelson2013}. To date, it has been widely assumed that the bias in hydrostatic mass is constant with redshift and mass, but it is unclear whether this assumption is valid. For upcoming cluster surveys, which will detect clusters out to $z\approx 1.5$, it is necessary to characterize the mass and redshift dependence of the mass bias and its impact on cosmological inferences.

Non-thermal pressure can also affects the interpretation of the angular power spectrum of the thermal SZ signal, originated from the inverse Compton scattering of the CMB photons off hot electrons in galaxy clusters. The amplitude of the angular power spectrum of the thermal SZ signal ($C_\ell$) is very sensitive to the amplitude of matter density fluctuations ($\sigma_8$) as $C_\ell \propto \sigma_8\,^{7-8}$ \citep{Komatsu_Seljak2002}. Non-thermal pressure is one of the main astrophysical uncertainties since most of the thermal SZ signal comes from integrated thermal pressure from the hot gas in the intracluster and intragroup medium at large radii, where the level of non-thermal pressure is comparable to that of thermal pressure, and where the energy injection from stars and active galactic nuclei are expected to be subdominant. The inclusion of non-thermal pressure support can change the amplitude of the thermal SZ power spectrum by as much as $60\%$ \citep{Shaw2010,Battaglia2010,Trac2011b}, significantly affecting its constraint on $\sigma_8$. Since the thermal SZ angular power spectrum gets contributions from galaxy groups and clusters in a wide range of redshifts and mass, a proper understanding of the mass and redshift dependence of the non-thermal pressure support is critical for using the SZ power spectrum and its high-order moment counterparts \citep{Bhattacharya2012,Hill2013} as cosmological probes. 

The upcoming {\em ASTRO-H} mission, equipped with high-resolution X-ray spectrometer, will measure internal gas motions in galaxy clusters from doppler broadening of emission lines \citep{Takahashi2010}. However, due to its limited sensitivity, the {\em ASTRO-H} measurements of the non-thermal pressure will be limited to only the inner regions of nearby massive clusters, and it will be difficult to extend these measurements to the outer regions or high-redshift clusters where the effects of non-thermal pressure are expected to be more significant.

In the absence of observational constraints, hydrodynamical cosmological simulations can serve as guides for characterizing the effects of non-thermal pressure, particularly at large cluster radii and at high redshifts. In this paper we build upon previous works \citep{Shaw2010,Battaglia2012a} by examining the non-thermal pressure fraction for a mass-limited sample of highly resolved massive galaxy clusters in a wide range of mass, redshifts and dynamical states. We show that the mean non-thermal pressure fraction as well as the gas velocity anisotropy profiles exhibit remarkable universality with redshift and mass, when the cluster mass is defined with respect to the mean mass density of the universe, instead of the critical density. We also find that these profiles show cluster-to-cluster scatter which depends primarily on the mass accretion rate of the clusters, which only affects the normalization of the profiles. We present fitting formulae for these universal profiles. These formulae should useful for characterizing the effects of non-thermal pressure on the hydrostatic mass bias, incorporating their effects in semi-analytic models of thermal SZ power spectrum, and calibrating analytical models of the non-thermal pressure profiles of clusters \citep[e.g.,][]{Shi2014}. 

The paper is organized as follows. In Section~\ref{sec:theory} we give an overview of the different mass definitions and describe our dynamical state proxy; in Section~\ref{sec:data} we describe our simulations of galaxy cluster formation; in Section~\ref{sec:results} we present our findings; and finally we offer our conclusions and discussions in Section~\ref{sec:summary}. 

\section{Theoretical Overview}
\label{sec:theory}

\subsection{Cluster Mass Definitions}
\label{sec:masses}

Galaxy clusters form at the intersections of large-scale filamentary structures in the universe.  As such, they have no well-defined physical edge, and we must adopt some convention for determining a boundary for the systems and their enclosed mass. The common approach is to define the boundary of a cluster as a sphere enclosing an average matter density equal to some chosen reference overdensity, $\Delta_{\rm ref}$, times some reference background density, $\rho_{\rm ref}$. The mass of the cluster is then
\begin{equation}
M_{\Delta_{\rm ref}} \equiv \frac{4\pi}{3}\Delta_{\rm ref} \rho_{\rm ref}(z) r_{\Delta_{\rm ref}}^3
\end{equation}
where $r_{\Delta_{\rm ref}}$ is the cluster radius. 
The two common choices of the background density $\rho_{\rm ref}$ are the critical density, $\rho_c(z)$, and the mean matter density, $\rho_m(z)$, 
\begin{align}
\rho_c(z) &= \frac{3H_0^2}{8\pi G}\left(\Omega_m(1+z)^3+\Omega_{\Lambda}\right), \\
\rho_m(z) &= \frac{3H_0^2}{8\pi G}\Omega_m(1+z)^3.
\end{align}
in the standard $\Lambda$CDM spatially flat cosmological model. The reference overdensity, $\Delta_{\rm ref}$, is usually chosen to be a number close to $18\pi^2 \approx 180$, correspond to the virial overdensity under the spherical collapse model in the Einstein-deSitter universe where $\Omega_m = 1- \Omega_\Lambda= 1$. In the more realistic flat $\Lambda$CDM universe, the virial overdensity is not constant and varies with redshift \citep[e.g.,][]{Bryan1998}. 

Conventionally $\rho_c(z)$ has been adopted as the reference background density for the cluster mass definition: $\rho_{\rm ref} = \rho_c(z)$, with $\Delta_{\rm ref} = \Delta_c = 500$ or $200$. This is a convenient choice because it depends only on the critical density and does not require additional prior knowledge of $\Omega_m$. $\Delta_c = 500$ has been most widely used in analyzing \emph{Chandra} and \emph{XMM-Newton} clusters, since it corresponds to the radius where these observatories can reliably measure gas density and temperature profiles of the intracluster medium (ICM). $\Delta_c = 200$ is also widely used since it is close to the virial overdensity in the spherical collapse model at $z=0$. 

A recent work by \citet{Diemer2014} demonstrates that while dark matter density profiles with halos defined with respect to the critical density $\rho_c(z)$ exhibit more self-similar behavior in their inner regions, adopting the reference density to be the mean matter density $\rho_{\rm ref} = \rho_m(z)$ results in a more self-similar density profile in the outskirts, where the clusters are more sensitive to the recent mass accretion. Since gas motions are driven by mass accretion and are more significant in the outer regions of clusters, the non-thermal pressure fraction profile is expected to be more self-similar when scaled with $r_{\Delta_m}$. In this paper, we compare the non-thermal pressure fraction profile as well as the gas velocity anisotropy profile using two different mass definitions based on $\Delta_c =200$ and $\Delta_m = 200$. For reference, $r_{500c} \approx 0.37\ r_{200m}$ and $r_{200c} \approx 0.58\ r_{200m}$ for our sample at $z=0$.

\subsection{Mass Accretion Rate}
\label{sec:dynamicalstate}

We use the mass accretion rate of galaxy clusters to identify their dynamical state at $z=0$. Following \cite{Diemer2014} we use the quantity $\Gamma$ as a proxy of mass accretion rate, such that
\begin{equation}
\Gamma_{200m} \equiv \Delta \log (M_{200m}(a))/ \Delta \log (a),  
\end{equation}
where $M_{200m}(a)$ is the mass of the cluster or its most massive progenitor measured at expansion factor $a$. $\Gamma$ is computed from the difference of each respective quantity between $z=0$ and $z=0.5$. The most massive progenitor of each cluster is tracked and identified using merger trees as described in \citet{Nelson2012}.  A higher $\Gamma$ means the halo experiences a greater {\em physical} mass accretion between $z=0$ and $z=0.5$.
We note that this definition of mass accretion naturally accounts for {\em physical} mass accretion and isolates effects of pseudo evolution in cluster mass \cite[due to evolution of the background reference density $\rho_{\rm ref}(z)$, see][]{Diemer2013a}, since we are comparing the increase in halo mass between two fixed redshifts.  

Also note that the choice of the redshift $z=0.5$ is arbitrary, and in fact $\Gamma$ is not the only (or necessarily the best) way of characterizing the mass accretion rate. We will explore alternative methods for quantifying the mass accretion rate in future work. For now, we adopt this definition to aid comparison between our results and the $N$-body results of \citet{Diemer2014}. 

\begin{figure*}[t]
\begin{center}
\includegraphics[scale=0.85]{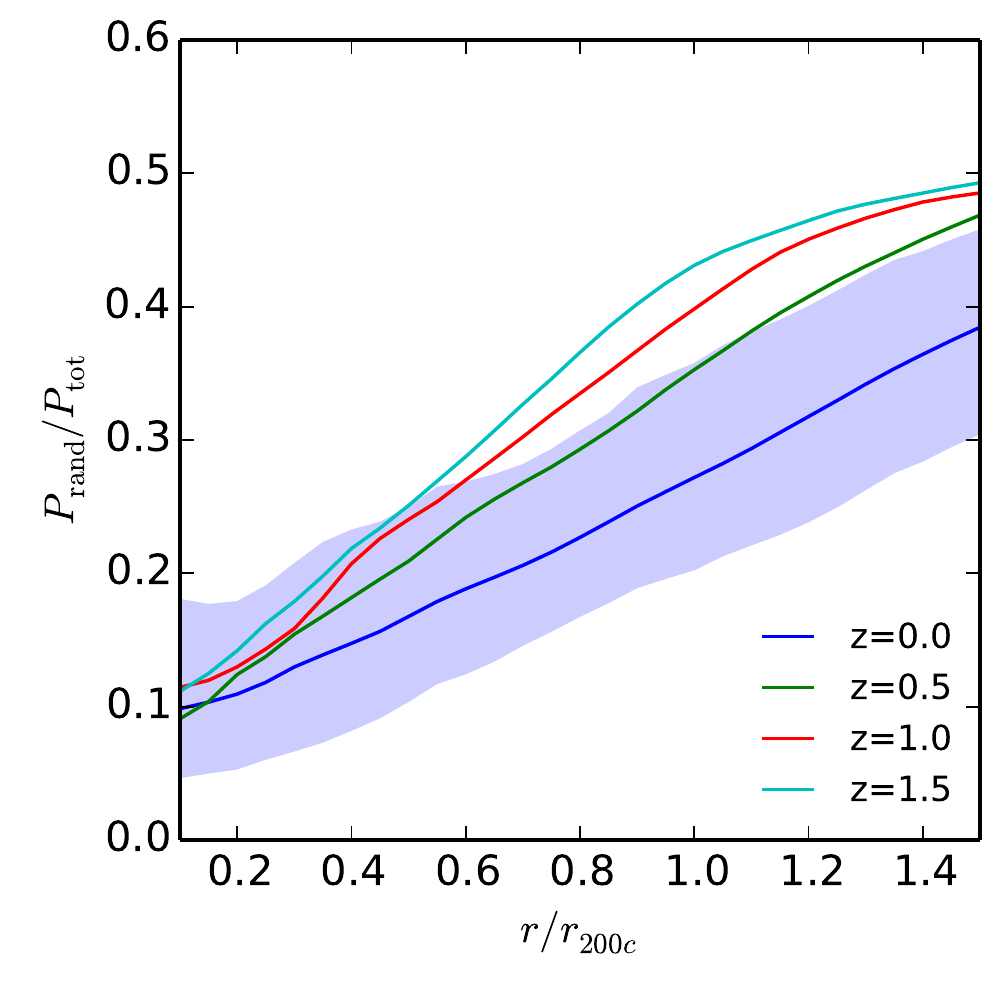}
\includegraphics[scale=0.85]{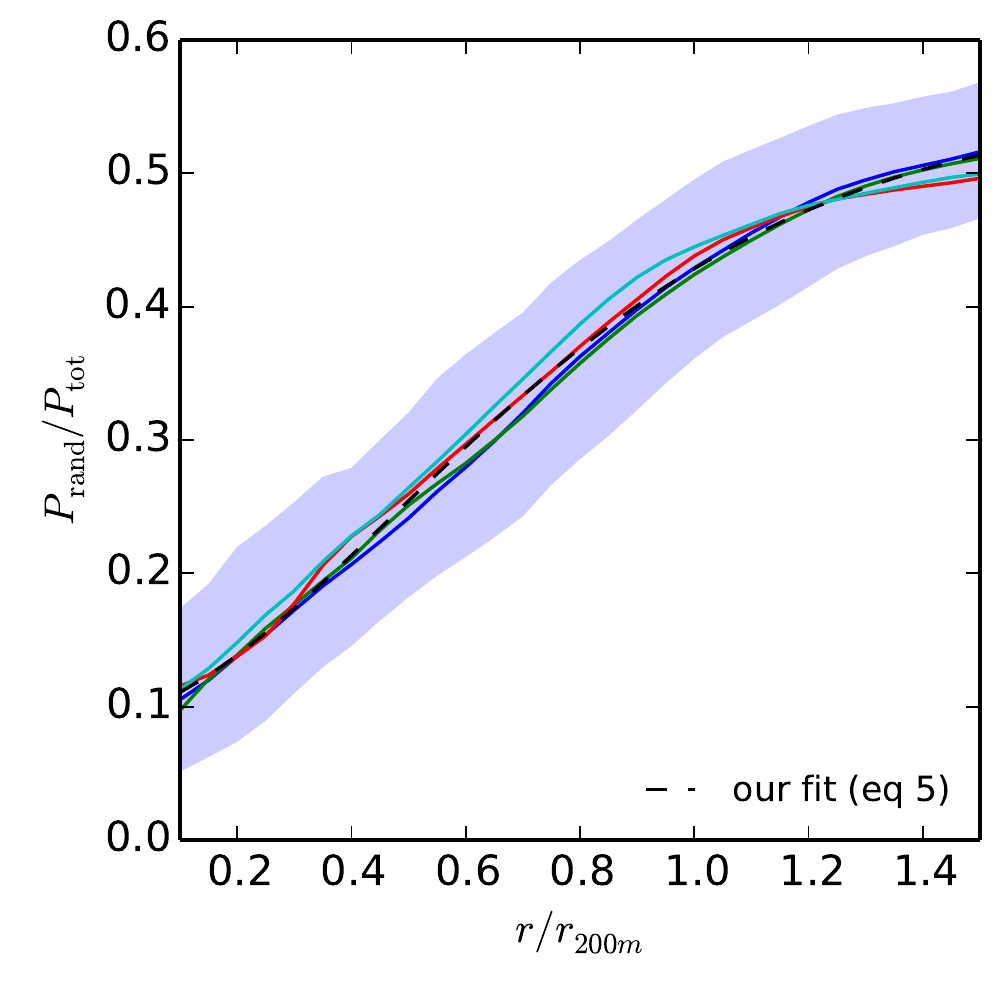}
\caption
{Redshift dependence of the profile of non-thermal pressure fraction $\pfrac$, with radius scaled with respect to $r_{200c}$ ({\em left}) and $r_{200m}$ ({\em right}). The shaded regions denote the 1-$\sigma$ scatter around mean at $z=0$. Our fitting formula is over plotted in the dashed line. }
\label{fig:p_z}
\end{center}
\end{figure*}

\section{Simulations}
\label{sec:data}

\subsection{Hydrodynamical Simulations of Galaxy Clusters}

We analyze simulated massive galaxy clusters presented previously in \cite{Nelson2013}. We refer the reader to that paper for more details. We briefly summarize the relevant parameters below.

In this work we analyze a high-resolution cosmological simulation of 65 galaxy clusters in a flat $\Lambda$CDM model with WMAP five-year ({\em WMAP5}) cosmological parameters: $\Omega_m = 1 - \Omega_{\Lambda} = 0.27$, $\Omega_b = 0.0469$, $h = 0.7$ and $\sigma_8 = 0.82$, where the Hubble constant is defined as $100h$~km~s$^{-1}$~Mpc$^{-1}$ and $\sigma_8$ is the mass variance within spheres of radius 8$h^{-1}$~Mpc. The simulation is performed using the Adaptive Refinement Tree (ART) $N$-body+gas-dynamics code \citep{Kravtsov1999, Kravtsov2002, Rudd2008}, which is an Eulerian code that uses adaptive refinement in space and time and non-adaptive refinement in mass \citep{Klypin2001} to achieve the dynamic ranges necessary  to resolve the cores of halos formed in self-consistent cosmological simulations. The simulation volume has a comoving box length of $500\,h^{-1}$~Mpc, resolved using a uniform $512^3$ grid and 8 levels of mesh refinement, implying a maximum comoving spatial resolution of $3.8\,h^{-1}$~kpc.  

Galaxy clusters are identified in the simulation using a variant of the method described in \citet{Tinker2008} (see \cite{Nelson2013} for a more detailed description of this method). We selected clusters with $M_{500c} \geq 3\times10^{14} h^{-1}M_{\odot}$ and  re-simulated the regions, defined as $5\times r_{\rm vir}$, surrounding the selected clusters with higher resolution. The resulting simulation has effective mass resolution of $2048^{3}$ surrounding the selected clusters, allowing a corresponding dark matter particle mass of $1.09 \times 10^9\, h^{-1}M_{\odot}$. The current simulation only models gravitational physics and non-radiative hydrodynamics. As shown in \citet{Lau2009}, the exclusion of cooling and star formation have negligible effect (less than a few percent) on the total contribution of gas motions to the non-thermal pressure support for $ r\geq 0.2 r_{500c}$, the radial range we focus on in this work. We show that our results are insensitive to dissipative physics in Appendix~\ref{sec:physics}.

To study the evolution of the non-thermal pressure fraction, we extract halos from four redshift outputs: $z=0.0, 0.5,1.0, 1.5$. At each redshift we apply an additional mass-cut to ensure mass-limited samples at all epochs. The mass-cuts and resulting sample sizes are as follows: 65 clusters with $M_{200m} \geq 6\times 10^{14} h^{-1}M_{\odot}$ at $z = 0$,  48 clusters with $M_{200m} \geq 2.5\times 10^{14} h^{-1}M_{\odot}$ at $z = 0.5$,  42 clusters with $M_{200m} \geq 1.3\times 10^{14} h^{-1}M_{\odot}$ at $z = 1.0$,  and 42 clusters with $M_{200m} \geq 7 \times 10^{13} h^{-1}M_{\odot}$ at $z = 1.5$.

\subsection{Computing the Non-thermal Pressure Fraction}

The non-thermal pressure fraction is defined as
\begin{equation}
\frac{P_{\rm rand}}{P_{\rm total}} = \frac{P_{\rm rand}}{P_{\rm rand}+P_{\rm therm}} = \frac{\sigma_{\rm gas}^2}{\sigma_{\rm gas}^2+(3kT/\mu m_p)}\label{eq:pfrac}
\end{equation}
where $k$ is the Boltzmann constant, $m_p$ is the proton mass, $\mu = 0.59$ is the mean molecular weight for the fully ionized ICM, $\sigma_{\rm gas}$ and $T$ are the mass-weighted 3D velocity dispersion and mass-weighted temperature of the gas averaged over spherical radial shells, respectively. 

To compute spherically averaged profiles of the mass-weighted 3D velocity dispersion and mass-weighted temperature we divide the analysis region for each cluster into 99 spherical logarithmic bins from $10\,h^{-1}{\rm kpc}$ to $10 \,h^{-1}{\rm Mpc}$ in the radial direction from the cluster center, which is defined as the position with the maximum binding energy. Our results are insensitive to the exact choice of binning.  We follow the procedure presented in \citet{Zhuravleva2013}, and we refer the reader to it for the details of the procedure and its impact on the non-thermal pressure measurements. Briefly, for each radial bin we exclude contribution from gas that lies in the high-density tail in the gas distribution, which remove small-scale fluctuations in the non-thermal pressure due to gas substructures while preserving the profiles of the global ICM. In addition, we smooth the profiles by applying the Savitzky-Golay filter used in \citet{Lau2009}.  

In computing the gas velocity dispersion, we first choose the rest frame of the system to be the center-of-mass velocity of the total mass interior to each radial bin. We rotate the coordinate system for each radial bin such that the $z$-axis aligns with the axis of the total gas angular momentum of that bin. We then compute mean $\langle v_i\rangle$ and mean-square gas velocities $\langle v^2_i \rangle$ weighted by the mass of each gas cell. The velocity dispersion is computed as $\sigma_i = \sqrt{\langle v^2_i\rangle-\langle{v}_i\rangle^2}$ for both the radial and tangential velocity components $\sigma_r$ and $\sigma_t$. The 3D velocity dispersion is simply $\sigma_{\rm gas} = \sqrt{(\sigma_r^2+\sigma_t^2)/3}$. 


\section{Results}
\label{sec:results}


\subsection{Universality with Redshift}
\label{sec:universality}

In Figure~\ref{fig:p_z} we present the non-thermal pressure fraction profile, $\pfrac(r)$, averaged for our cluster sample at four redshifts $z=0.0,0.5,1.0,1.5$. In the two panels we show the evolution of the profiles normalized using two different halo radii $r_{200c}$ and $r_{200m}$ on the left and right, respectively. The shaded regions depict the 1-$\sigma$ scatter around the mean for $z=0$. The scatter is comparable at all redshifts and has been omitted for clarity. For $r_{200c}$, there is a strong redshift evolution of both the shape and normalization of $\pfrac$. At low redshift, the non-thermal pressure fraction is $\approx 10\%$ in the inner regions of the clusters, increasing up to $> 40\%$ outside of $r>r_{200c}$. At higher redshift, the non-thermal pressure systematically constitutes a larger fraction of the total pressure support in the systems. Moreover, the profile increases with radius more rapidly, reaching up to $\approx 45\%$ of the total pressure at $r_{200c}$ at $z = 1.5$, twice the fraction at $z = 0$. 

In the right panel, we again show the redshift evolution of $\pfrac$, however, this time normalizing the profiles using the halo radius $r_{200m}$. When the halo radius is defined with the mean background density of the universe, the radial dependence of $\pfrac$ remains, with $\approx 10\%$ non-thermal pressure support in the core rising steadily to $r=r_{200m}$ where the pressure fraction flattens out to $\approx 50\%$ pressure support in the cluster outskirts. However, the very strong redshift dependence seen using $r_{200c}$ as the halo radius has completely disappeared. The universality we find in our non-thermal pressure fraction is in agreement with the very weak redshift dependence \citet{Battaglia2012a} show in their non-thermal pressure fraction profile, also scaled with $r_{200m}$.

The simple reason behind this universality with $r_{200m}$ is that the non-thermal pressure fraction is sensitive to the mass accretion rate of the clusters. The rate at which mass is accreting depends on the physical density contrast between the halo and the mean background density, precisely the quantity used in the definition of $r_{200m}$.  We will provide a more detailed explanation of this redshift universality in our follow-up paper. 

\subsection{Dependence on Mass and Mass Accretion Rate}
\label{mass_accretion_rate}

\begin{figure}[t]
\begin{center}
\includegraphics[scale=0.85]{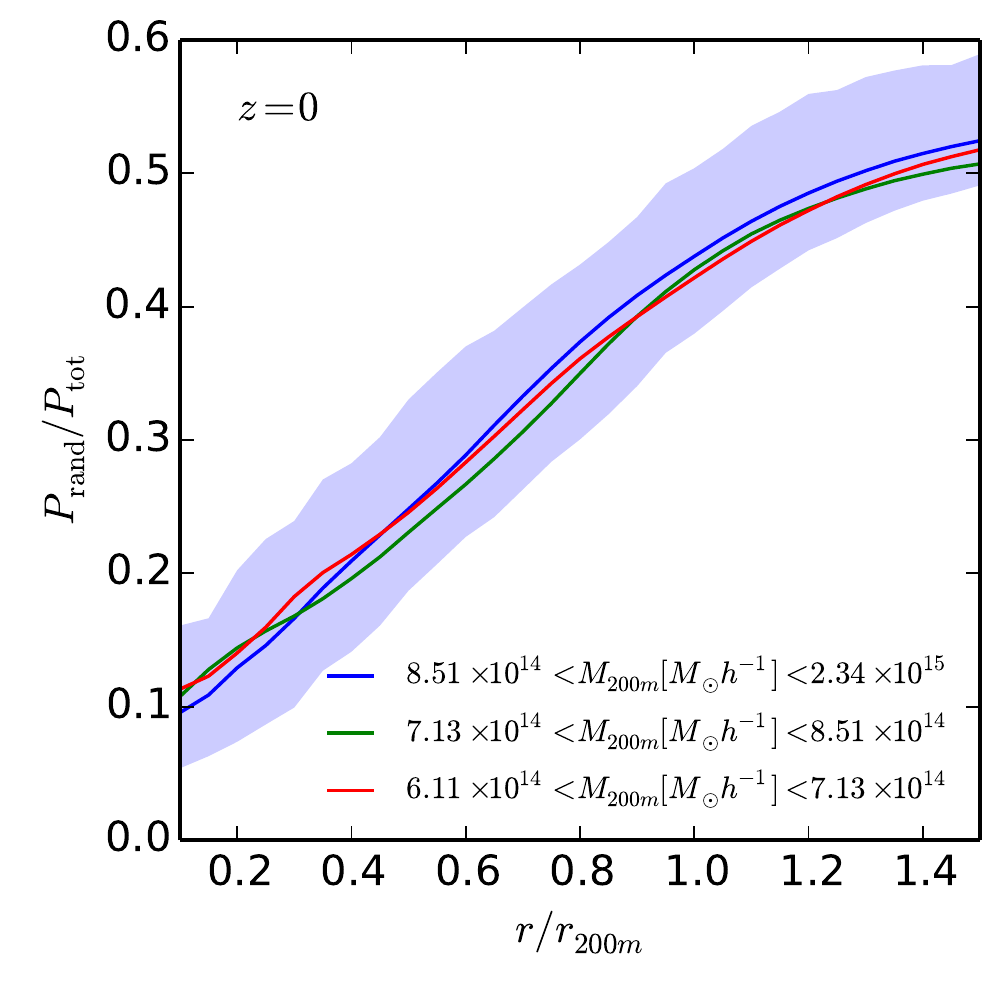}
\caption{Mass dependence of $\pfrac$ profile at $ z = 0$. The sample has been divided into three equal sized mass bins as denoted in the legend, shown in blue, green and red lines from most massive to least massive subsamples. The shaded regions denote the 1-$\sigma$ scatter around the mean in the most massive bin.}
\label{fig:p_mass}
\end{center}
\end{figure}

In order to characterize the scatter in the non-thermal pressure fraction profile, we examine the mass and dynamical state dependence of the profile.  In Figure~\ref{fig:p_mass}, we divide the sample into three equal sized mass bins at $z$ = 0, shown in red, green and blue from lowest mass to highest mass subsamples, respectively. We find no systematic dependence on mass in the profile. However, it is worth noting that our sample only encompasses the massive end of the cluster population with $M_{200m} \geq 6\times 10^{14}h^{-1}M_{\odot}$ or $M_{200c} \geq 4\times 10^{14}h^{-1}M_{\odot}$ correspondingly.

\begin{figure*}[t]
\begin{center}
\includegraphics[scale=0.85]{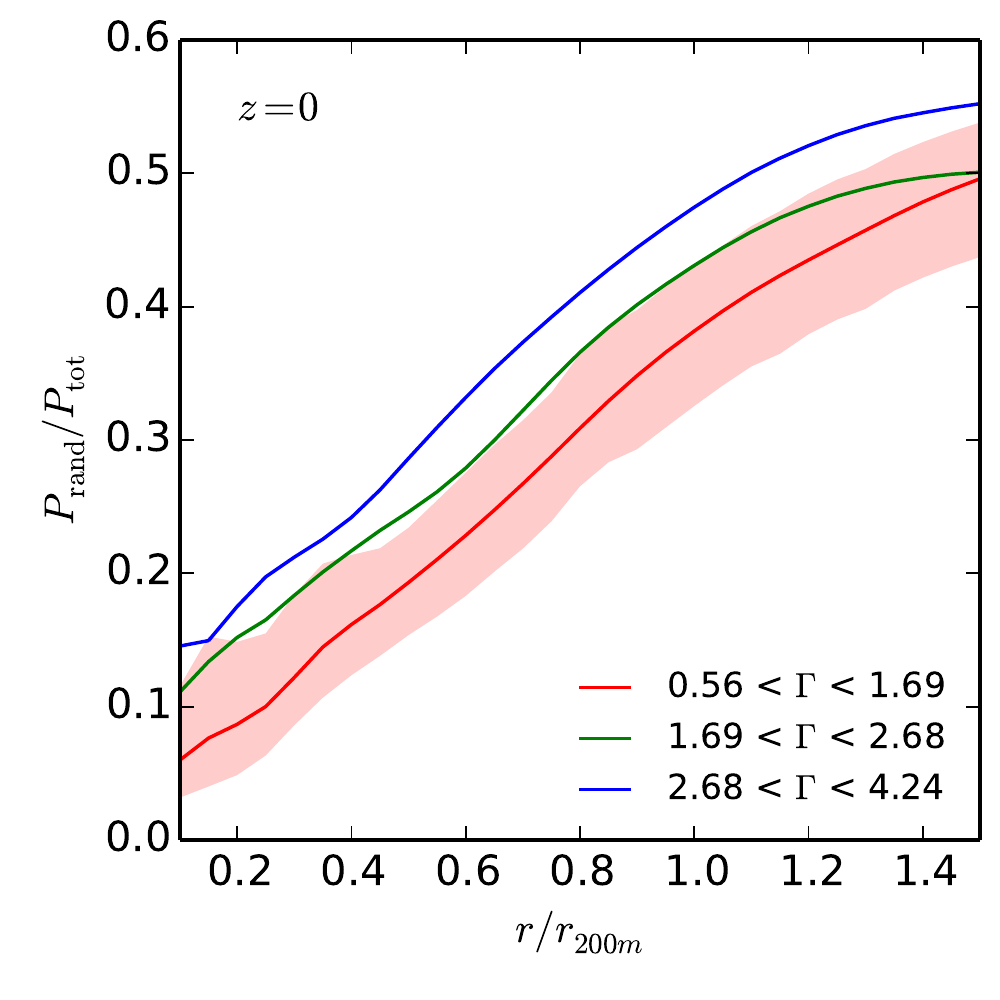}
\includegraphics[scale=0.85]{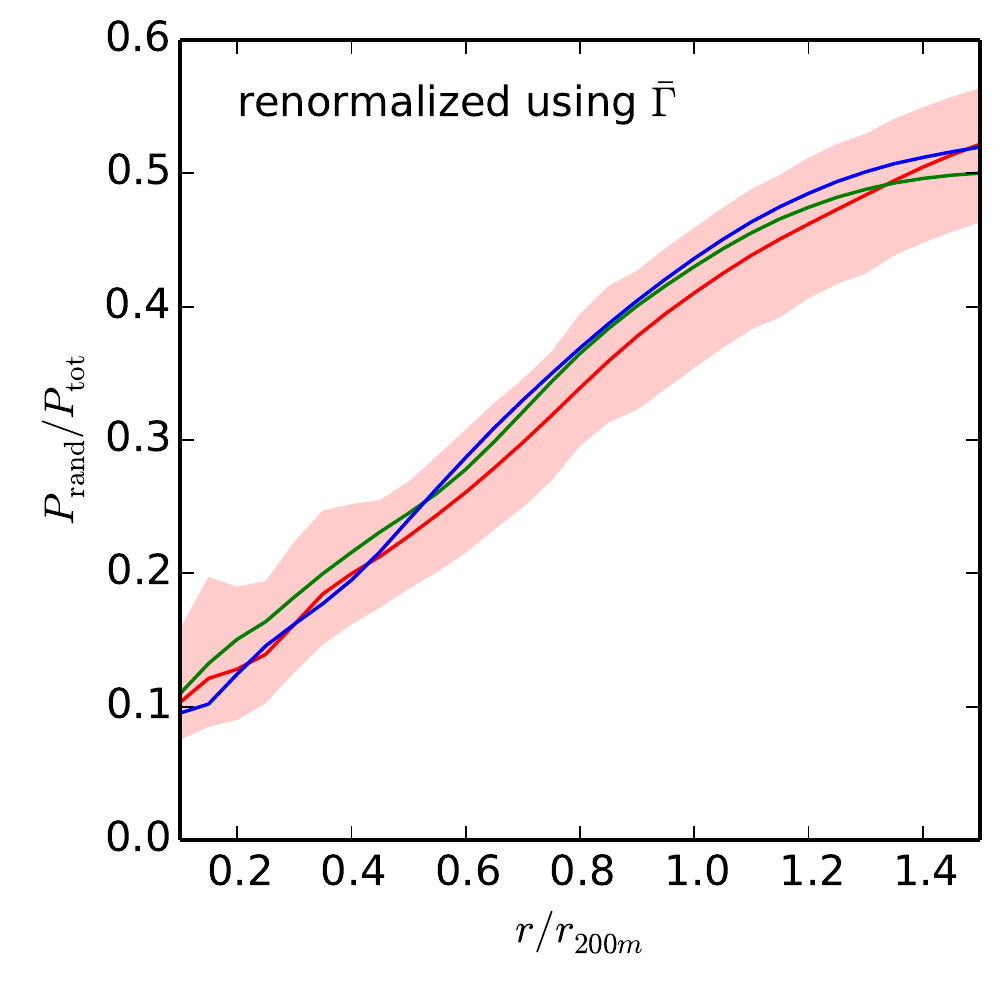}
\caption
{Dynamical state dependence of the $\pfrac$ profiles at $z=0$. The sample has been divided into three equal sized mass accretion rate bins as denoted in the legend, shown in red, green and blue from most slowly to most rapidly accreting clusters. In the right panel the profiles have been renormalized by the ratio of Eq~\ref{eq:fit} to Eq~\ref{eq:gamma} for the mean $\Gamma$ value in each bin, respectively, in order to remove the dynamical state dependence. The shaded regions denote the 1-$\sigma$ scatter in the most relaxed bin.}
\label{fig:p_accretion}
\end{center}
\end{figure*}

In the left panel of Figure~\ref{fig:p_accretion}, we investigate the dynamical state dependence of the non-thermal pressure fraction at $z= 0$.  The mean profiles are shown for three equal sized $\Gamma$ bins, shown in red, green, and blue lines from the most slowly to most rapidly accreting systems, respectively. The high $\Gamma$ subsample has systematically higher non-thermal pressure fraction at all radii by $5\%-15\%$ than the most slowly accreting subsample. This is consistent with previous works which also find significantly larger fractions of non-thermal pressure in merging or unrelaxed systems \citep{Vazza2011, Nelson2012}.  Given the small mass dependence seen in Figure~\ref{fig:p_mass}, the small amount of scatter in the $\pfrac$ profile in Figure~\ref{fig:p_z} is likely dominated by the range of dynamical states of the clusters in the sample. Note, however, that the main differences are in the amplitude of the non-thermal pressure fraction, and we see very little change in the slope of the profile.

\subsection{Velocity Anisotropy}
\label{sec:anisotropy}

\begin{figure*}[t]
\begin{center}
\includegraphics[scale=0.85]{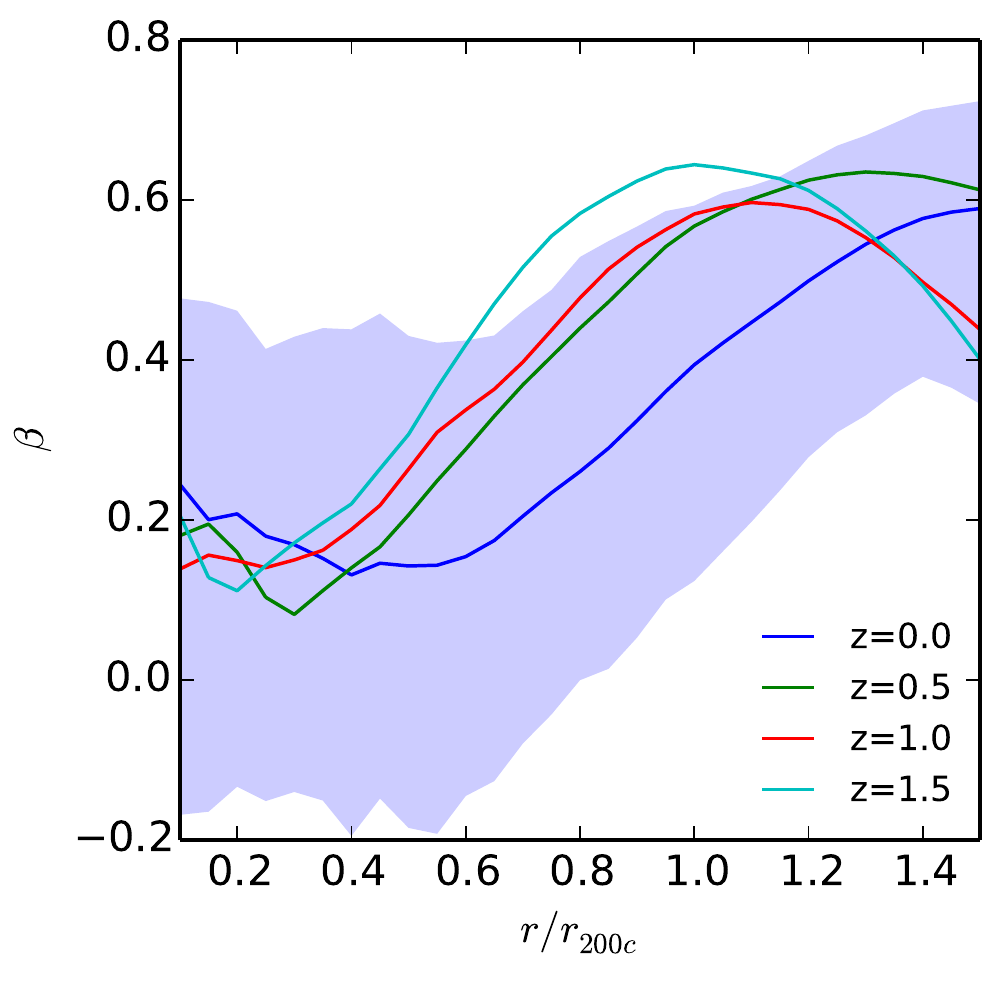}
\includegraphics[scale=0.85]{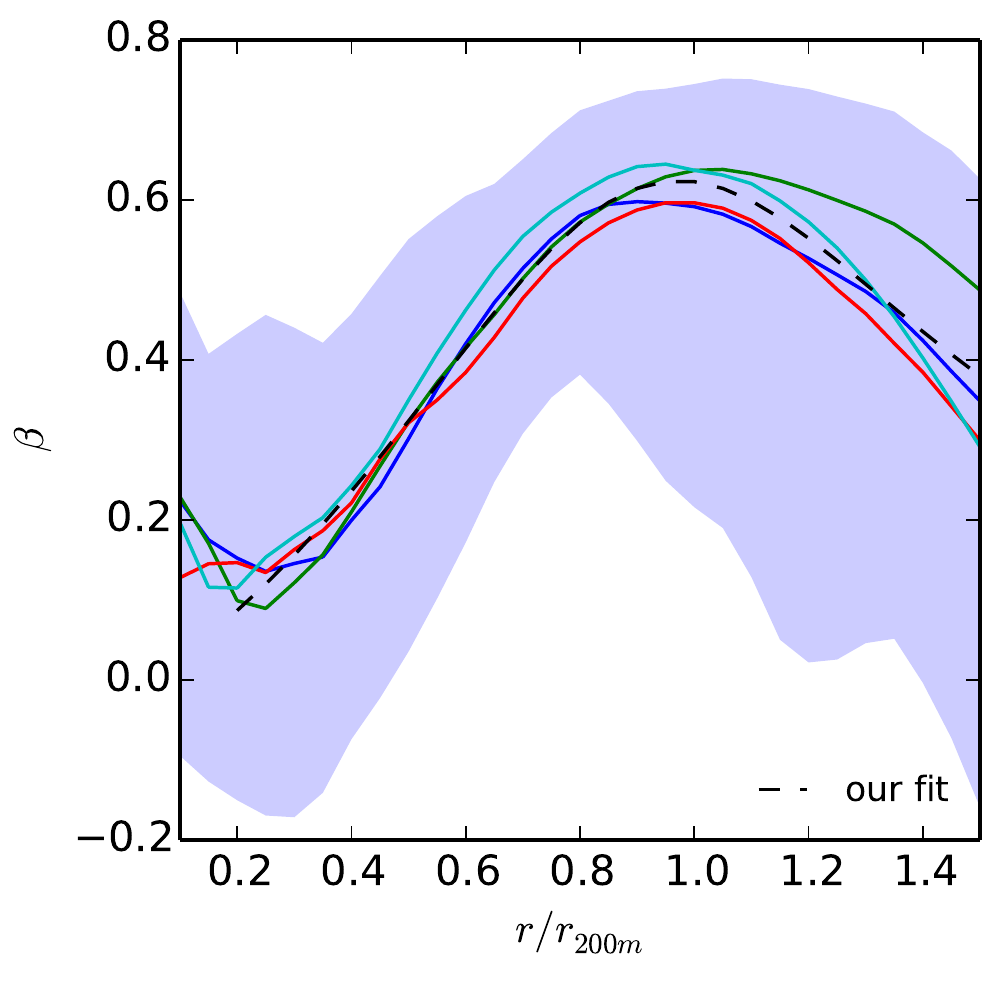}
\caption
{Redshift dependence of the profile of velocity anisotropy, with radius scaled with respect to $r_{200c}$ ({\em left}) and $r_{200m}$ ({\em right}). The shaded regions denote the 1-$\sigma$ scatter around the mean at $z=0$.}
\label{fig:anisotropy}
\end{center}
\end{figure*}

In our definition of $P_{\rm rand}$ in Equation~(\ref{eq:pfrac}), we assume the gas velocities in the galaxy clusters are isotropic. In this section we explore the relative importance of the radial and tangential components in the non-thermal pressure fraction by examining the velocity anisotropy parameter $\beta$,
\begin{equation}\label{eq:anisotropy}
\beta (r) = 1 - \frac{\sigma_t^2(r)}{2\sigma_r^2(r)},
\end{equation}
where a positive (negative) value of $\beta$ indicates more radial (tangential) motion. 
In Figure~\ref{fig:anisotropy}, we examine the redshift evolution of the anisotropy with the cluster radius scaled with respect to $r_{200c}$ ({\em left} panel) and $r_{200m}$ ({\em right} panel). Similar to the $\pfrac$ profile, the profile scaled with respect to critical density shows significant redshift dependence. The profile scaled with respect to mean matter density, on the other hand, shows universality across the radial range of $ 0.3 \lesssim r/r_{200m} \lesssim 1.5$.  The gas velocities are slightly radial in the inner regions (e.g., $\beta \approx 0.1$ at $r = 0.2 r_{200m}$) and becomes increasingly radial, reaching $\beta \approx 0.6$ around $r=r_{200m}$. 

In the cluster outskirts, the anisotropy decreases as the gas motions become more isotropic again. The apparent drop in $\beta$ is partly due to our definition of the velocity anisotropy in Equation~(\ref{eq:anisotropy}), in which we neglected the contributions from coherent motions in both radial and tangential directions. At large radii, there is a coherent radial motion toward the cluster center \citep[e.g., see Figure~1 in][]{Lau2009}, causing $\sigma_r$ (representing the random radial motions) and hence $\beta$ to decrease with radius. 

\subsection{Fitting Formulae}
\label{function}

In this section, we provide fitting formulae for the universal non-thermal pressure fraction and gas velocity anisotropy profiles discussed in the previous sections. By using $r_{200m}$ for the halo radius, our universal non-thermal pressure profile is well-described by the following fitting formula,
\begin{equation}
\frac{P_{\rm rand}}{P_{\rm total}} (r) = 1-A \left\{1+\exp\left[-\left(\frac{r/r_{200m}}{B}\right)^{\gamma}\right]\right\}
\label{eq:fit}
\end{equation}
where the best fit parameters are $A = 0.452 \pm 0.001$, $B = 0.841 \pm 0.008$, and $\gamma = 1.628 \pm 0.019$. This formula provides a good description of the mean profile with the accuracy of about $10\%$ in the radial range of $ 0.1 \leq r/r_{200m} \leq 1.5$ out to $z = 1.5$. The best fit line is plotted in the right panel of Figure~\ref{fig:p_z}. In the \hyperref[sec:appendix]{Appendix}, we also supply the adjusted fitting formula for the scaling of radii with respect to critical. Despite using $r_{\Delta c}$, we are still able to provide an accurate fit out to $z=1$, since the formula is based on the universal $r_{\Delta m}$ profile.

As shown in the left panel of Figure~\ref{fig:p_accretion}, the varied mass accretion histories of the clusters in our sample is a major source of scatter in the non-thermal pressure fraction profile. To account for the mass accretion rate, we provide a correction factor for the normalization of our fitting formula at $z=0$ using $\Gamma$ as a parameter,
\begin{align}
&\frac{P_{\rm rand}}{P_{\rm total}} (r; z=0)  \nonumber \\
& = 1 - (0.509- 0.026 \Gamma_{200m})\ \left\{1+\exp\left[-\left(\frac{r/r_{200m}}{B}\right)^{\gamma}\right]\right\},
\label{eq:gamma}
\end{align}
where we keep our best fit parameters $B= 0.841$ and $\gamma = 1.628$ obtained earlier by fitting in the radial range of $ 0.1 \leq r/r_{200m} \leq 1.5$. We use the best fit $A$ (Equation~(\ref{eq:fit})) values for the profiles of each of the 65 clusters in our sample, with $B$ and $\gamma$ fixed at our best fit values, to determine the function of $\Gamma$. In the right panel of Figure~\ref{fig:p_accretion}, we show the same three $\Gamma$ bins as in the left panel but renormalized using Equation~(\ref{eq:gamma}), fit using the mean $\Gamma$ value in each bin respectively. While there is some residual variation in the slope of the profile for different mass accretion bins, including this normalization correction factor decreases the scatter at $r_{200m}$ from $0.067$ to $0.053$, leading to 1-$\sigma$ agreement between the adjusted profiles.

Previous attempts to characterize the non-thermal pressure fraction have used the $r_{200c}$ (or $r_{500c}$) as the cluster radius and therefore have included an additional factors to account for its strong redshift evolution \citep{Shaw2010} and mass dependence \citep{Battaglia2012a}. We have confirmed that the $r_{200c}$ profiles for our data are well fit by their fitting formulae at $z=0.0$ in Appendix~\ref{sec:scaling}. 

Lastly, we provide a fitting formula for the universal gas velocity anisotropy profile shown in the right panel of Figure~\ref{fig:anisotropy}. We adopt the following fitting formula,
\begin{equation}
\beta (r) = \frac{(r/r_t)^{-a}}{(1 + (r/r_t)^{b})^{c/b}},
\label{eq:fit_beta}
\end{equation}

fit between $0.2 \leq r/r_{200m} \leq 1.5$, where the best fit parameters are $r_t = 1.083 \pm 0.028$, $a = 2.643 \pm 0.211$,  $b = -5.637 \pm 0.183$ and $c = -4.090 \pm 0.169$. The profile is well fit with the accuracy of about 20\% between $0.3 \leq r/r_{200m} \leq 1.3$ out to $z = 1.5$, with the exception of $z=0.5$, which is only a robust fit for $r \leq r_{200m}$.

\section{Summary and Discussion}
\label{sec:summary}

In this work we presented the redshift and mass independent non-thermal pressure fraction profile using a mass-limited, cosmologically representative sample of 65 massive galaxy clusters from a high resolution hydrodynamical cosmological simulation. This result is relevant in accounting for the systematic effects of non-thermal pressure on X-ray and microwave measurements of galaxy clusters and cosmological inferences based on these measurements.

We found that the mean non-thermal pressure fraction profile exhibits remarkable universality in redshift and mass when we define the size of cluster halos using the mean matter density of the universe, instead of the critical density.  However, we also showed that there is strong dependence in the non-thermal pressure fraction profile on the halo's mass accretion rate: clusters that are rapidly accreting have an overall higher non-thermal pressure fraction. As such, the mass accretion rate is a major source of systematic scatter in the mean non-thermal pressure fraction profile. 

A robust and quantitative proxy for measuring mass accretion rate is therefore needed to account for this effect, especially with the upcoming multi-wavelength cluster surveys where statistical errors will be considerably smaller than systematic uncertainties arising from our ignorance of cluster astrophysics. We note that the current method of characterizing the mass accretion rate using the fractional mass increase between $z=0$ and $z=0.5$ is by no means unique, and can only be applied to $z=0$ clusters. Future work should focus on developing quantitative measures of the mass accretion rate of halos that can be applied to halos across a wide range of redshifts, and relate these measures to observable proxies of the dynamical states of clusters.

We found no systematic mass dependence in the universal non-thermal pressure fraction profile. But given that our sample contains only massive clusters, the independence in mass should be checked with sample of lower mass halos.  Since slowly accreting halos have smaller non-thermal pressure fraction, we expect that lower mass groups and galaxies, which should experience less physical mass accretion than high mass clusters, to have lower non-thermal pressure fraction profile. However, we note that smaller mass halos are more susceptible to non-gravitational physics (such as gas cooling and energy injections from stars and active galactic nuclei) which can influence the net accretion rate into and within the halos in a non-trivial way. 

%
%

We found that the gas velocity is predominantly radial, with the velocity anisotropy parameter increasing from $ \approx 0.1$ to $ \approx 0.6 $ from $ 0.2 r_{200m} $ to $r_{200m}$. Moreover, we found that the velocity anisotropy profile is also universal across redshift when halos are defined using the mean matter density, indicating that gas velocity anisotropy is also a self-similar quantity. This result can be useful since the velocity anisotropy cannot be easily measured in observations, as we can only measure line-of-sight velocities from Doppler width measurements, e.g., with the upcoming {\em ASTRO-H} in the near future. Measurements of gas motions tangential to the line-of-sight are possible with resonant scattering but difficult \citep[e.g.,][]{Zhuravleva2011}. 

We provided fitting formulae for the universal non-thermal pressure fraction and gas velocity anisotropy profiles that work remarkably well within $r_{200m}$ and out to redshift $z=1.5$.  One application of our fitting formulas is the recovery of total mass of relaxed clusters by accounting for the hydrostatic mass bias. The effect of velocity anisotropy should be included in the mass recovery, since the formalism for the full mass recovery depends on the relative contribution of the radial and tangential components of the non-thermal pressure \citep{Rasia2004,Lau2009}. We note that while non-thermal pressure motions due to random gas motions are not the only contribution to the hydrostatic mass bias, the other mass terms due to coherent gas rotation and radial gas acceleration contribute typically only a few percent to the total mass bias, and as such are subdominant to the non-thermal pressure due to gas motions \citep{Suto2013,Lau2013}.  Another application of the fitting formula is the assessment of systematic uncertainties in the thermal SZ power spectrum due to non-thermal pressure. The universal, redshift independent profile provided here should make the implementation of non-thermal pressure support in the modeling of the thermal SZ power spectrum robust and straightforward. 

While our current simulation does not include radiative cooling, star formation or energy feedback from stars and/or active galactic nuclei, we have examined the effect of these additional physics on the non-thermal pressure fraction and gas velocity anisotropy in group and cluster size halos taken from \citet{Nagai2007a} and found no systematic dependence on gas physics in the radial range of $0.1 \lesssim r/r_{200m} \lesssim 1.5$.  We note, however, that our simulation does not model plasma effects which can amplify gas turbulence and provide extra non-thermal pressure support \citep[e.g.,][]{Parrish2012}. Physical viscosity, on the other hand, can decrease the level of gas turbulence which lowers the non-thermal pressure support. The results presented in this paper based on hydrodynamical simulations serve as baseline for further studies of these effects. Magnetic fields and cosmic rays can also provide additional non-thermal pressure \citep[e.g.,][]{Lagana2010}.  However, their contributions are expected to be small. The typical magnetic field strength of $\lesssim 10 \mu\mathrm{G}$ in the ICM corresponds to magnetic pressure fraction of $\lesssim 1\%$. Similarly, the ratio of the cosmic ray pressure to total pressure is constrained to $\lesssim 1\%$, set by the $\gamma$-ray observations of {\em Fermi}-LAT \citep{Fermi2013}. It is important to note, however, that the constraints on the contribution from cosmic rays assume that the cosmic ray distribution follows that of the thermal ICM and, therefore, a flattened distribution of cosmic rays could result in an increased contribution to the ICM energy \citep[e.g.,][]{Zandanel2014}

The upcoming {\em ASTRO-H} mission will measure gas motions in galaxy clusters and should provide observational constraints on the level of the non-thermal pressure fraction in these systems. However, the observational constraints will be limited to inner regions ($\lesssim r_{2500c}\approx 0.2\,r_{200m}$) of nearby massive clusters, due to the lack of sensitivities in low-density regions in cluster outskirts.  Extending these measurements to the outskirts or high-redshift clusters must await the next generation of X-ray missions, such as \emph{SMART-X}\footnote{\url{http://smart-x.cfa.harvard.edu/}} and/or \emph{Athena+}\footnote{\url{http://athena2.irap.omp.eu/}}. Alternatively, kinematic SZ effect can probe internal gas motions of electrons in galaxy clusters \citep{Inogamov2003,Nagai2003}.  Since the SZ signal is independent of redshift and linearly proportional to gas density (unlike X-ray emission which is proportional to gas density squared), measurements of the kSZ effect with high-resolution, multifrequency radio telescopes, such as CCAT\footnote{\url{http://www.ccatobservatory.org}}, might enable characterization of the non-thermal pressure in the outer regions of high-redshift clusters. 

Previous works have used the definition of cluster mass normalized with respect to the critical density of the universe.  In this work, we argue that an alternative definition based on the mean mass density of the universe is a preferred choice for the non-thermal pressure profile as well as velocity anisotropy of gas in clusters. It would be interesting to check whether other gas properties exhibit similar universality when the cluster profiles are normalized with respect to the mean mass density. We will investigate these issues in our next paper and explore their implications for understanding the evolution cluster gas structure and their application to cosmology. 

\acknowledgments 
We thank Nick Battaglia, Eiichiro Komatsu, and the anonymous referee for comments on the manuscript. This work was supported in part by NSF grant AST-1009811, NASA ATP grant NNX11AE07G, NASA Chandra grants GO213004B and TM4-15007X, the Research Corporation, and by the facilities and staff of the Yale University Faculty of Arts and Sciences High Performance Computing Center. 

\bibliography{ms}

\begin{thebibliography}{}
\expandafter\ifx\csname natexlab\endcsname\relax\def\natexlab#1{#1}\fi

\bibitem[{{Allen} {et~al.}(2011){Allen}, {Evrard}, \& {Mantz}}]{Allen2011}
{Allen}, S.~W., {Evrard}, A.~E., \& {Mantz}, A.~B. 2011,
  \href{http://dx.doi.org/10.1146/annurev-astro-081710-102514}{\araa},
  \href{http://adsabs.harvard.edu/abs/2011ARA%26A..49..409A}{49},
  \href{http://adsabs.harvard.edu/abs/2011ARA%26A..49..409A}{409}

\bibitem[{{Applegate} {et~al.}(2014){Applegate}, {von der Linden}, {Kelly},
  {Allen}, {Allen}, {Burchat}, {Burke}, {Ebeling}, {Mantz}, \&
  {Morris}}]{Applegate2014}
{Applegate}, D.~E., {von der Linden}, A., {Kelly}, P.~L., {et~al.} 2014,
  \href{http://dx.doi.org/10.1093/mnras/stt2129}{\mnras},
  \href{http://adsabs.harvard.edu/abs/2014MNRAS.439...48A}{439},
  \href{http://adsabs.harvard.edu/abs/2014MNRAS.439...48A}{48}

\bibitem[{Battaglia {et~al.}(2012)Battaglia, Bond, Pfrommer, \&
  Sievers}]{Battaglia2012a}
Battaglia, N., Bond, J.~R., Pfrommer, C., \& Sievers, J.~L. 2012,
  \href{http://dx.doi.org/10.1088/0004-637X/758/2/74}{\apj},
  \href{http://arxiv.org/abs/1109.3709
  http://stacks.iop.org/0004-637X/758/i=2/a=74?key=crossref.04717e7b9fbc682f2ed4f409f671f86f}{758},
  \href{http://arxiv.org/abs/1109.3709
  http://stacks.iop.org/0004-637X/758/i=2/a=74?key=crossref.04717e7b9fbc682f2ed4f409f671f86f}{74}

\bibitem[{Battaglia {et~al.}(2010)Battaglia, Bond, Pfrommer, Sievers, \&
  Sijacki}]{Battaglia2010}
Battaglia, N., Bond, J.~R., Pfrommer, C., Sievers, J.~L., \& Sijacki, D. 2010,
  \href{http://dx.doi.org/10.1088/0004-637X/725/1/91}{\apj},
  \href{http://stacks.iop.org/0004-637X/725/i=1/a=91?key=crossref.2df4b2e9a9e84a7f20b5ddfe7690c8f4}{725},
  \href{http://stacks.iop.org/0004-637X/725/i=1/a=91?key=crossref.2df4b2e9a9e84a7f20b5ddfe7690c8f4}{91}

\bibitem[{Bhattacharya {et~al.}(2013)Bhattacharya, Habib, Heitmann, \&
  Vikhlinin}]{Bhattacharya2011}
Bhattacharya, S., Habib, S., Heitmann, K., \& Vikhlinin, A. 2013,
  \href{http://dx.doi.org/10.1088/0004-637X/766/1/32}{\apj},
  \href{http://arxiv.org/abs/1112.5479
  http://stacks.iop.org/0004-637X/766/i=1/a=32?key=crossref.f55c61babcbb7ac327ea6c8508aed69d}{766},
  \href{http://arxiv.org/abs/1112.5479
  http://stacks.iop.org/0004-637X/766/i=1/a=32?key=crossref.f55c61babcbb7ac327ea6c8508aed69d}{32}

\bibitem[{{Bhattacharya} {et~al.}(2012){Bhattacharya}, {Nagai}, {Shaw},
  {Crawford}, \& {Holder}}]{Bhattacharya2012}
{Bhattacharya}, S., {Nagai}, D., {Shaw}, L., {Crawford}, T., \& {Holder}, G.~P.
  2012, \href{http://dx.doi.org/10.1088/0004-637X/760/1/5}{\apj},
  \href{http://adsabs.harvard.edu/abs/2012ApJ...760....5B}{760},
  \href{http://adsabs.harvard.edu/abs/2012ApJ...760....5B}{5}

\bibitem[{Bryan \& Norman(1998)}]{Bryan1998}
Bryan, G.~L., \& Norman, M.~L. 1998,
  \href{http://dx.doi.org/10.1086/305262}{\apj},
  \href{http://iopscience.iop.org/0004-637X/495/1/80}{1},
  \href{http://iopscience.iop.org/0004-637X/495/1/80}{80}

\bibitem[{Diemer \& Kravtsov(2014)}]{Diemer2014}
Diemer, B., \& Kravtsov, A.~V. 2014, \apj, in press,
  arXiv:\href{http://adsabs.harvard.edu/abs/2014arXiv1401.1216D}{1401.1216}

\bibitem[{Diemer {et~al.}(2013)Diemer, More, \& Kravtsov}]{Diemer2013a}
Diemer, B., More, S., \& Kravtsov, A.~V. 2013,
  \href{http://dx.doi.org/10.1088/0004-637X/766/1/25}{\apj},
  \href{http://arxiv.org/abs/1207.0816
  http://stacks.iop.org/0004-637X/766/i=1/a=25?key=crossref.0ca1347647fed29c8a38dfd0995d23a0}{766},
  \href{http://arxiv.org/abs/1207.0816
  http://stacks.iop.org/0004-637X/766/i=1/a=25?key=crossref.0ca1347647fed29c8a38dfd0995d23a0}{25}

\bibitem[{{{\em Fermi}-LAT Collaboration}(2013)}]{Fermi2013}
{{\em Fermi}-LAT Collaboration}. 2013, \apj, in press,
  arXiv:\href{http://adsabs.harvard.edu/abs/2013arXiv1308.5654T}{1308.5654}

\bibitem[{Evrard {et~al.}(1996)Evrard, Metzler, \& Navarro}]{Evrard1996}
Evrard, A.~E., Metzler, C.~A., \& Navarro, J.~F. 1996,
  \href{http://dx.doi.org/10.1086/177798}{\apj},
  \href{http://adsabs.harvard.edu/doi/10.1086/177798}{469},
  \href{http://adsabs.harvard.edu/doi/10.1086/177798}{494}

\bibitem[{{Hill} \& {Sherwin}(2013)}]{Hill2013}
{Hill}, J.~C., \& {Sherwin}, B.~D. 2013,
  \href{http://dx.doi.org/10.1103/PhysRevD.87.023527}{Phys. Rev. D},
  \href{http://adsabs.harvard.edu/abs/2013PhRvD..87b3527H}{87},
  \href{http://adsabs.harvard.edu/abs/2013PhRvD..87b3527H}{023527}

\bibitem[{Hu \& Kravtsov(2003)}]{Hu2003}
Hu, W., \& Kravtsov, A.~V. 2003, \href{http://dx.doi.org/10.1086/345846}{\apj},
  \href{http://stacks.iop.org/0004-637X/584/i=2/a=702}{584},
  \href{http://stacks.iop.org/0004-637X/584/i=2/a=702}{702}

\bibitem[{Inogamov \& Sunyaev(2003)}]{Inogamov2003}
Inogamov, N.~A., \& Sunyaev, R.~A. 2003,
  \href{http://dx.doi.org/10.1134/1.1631412}{Astron. Lett.},
  \href{http://arxiv.org/abs/astro-ph/0310737
  http://link.springer.com/10.1134/1.1631412}{29},
  \href{http://arxiv.org/abs/astro-ph/0310737
  http://link.springer.com/10.1134/1.1631412}{791}

\bibitem[{Klypin {et~al.}(2001)Klypin, Kravtsov, Bullock, \&
  Primack}]{Klypin2001}
Klypin, A., Kravtsov, A.~V., Bullock, J.~S., \& Primack, J.~R. 2001,
  \href{http://dx.doi.org/10.1086/321400}{\apj},
  \href{http://stacks.iop.org/0004-637X/554/i=2/a=903}{554},
  \href{http://stacks.iop.org/0004-637X/554/i=2/a=903}{903}

\bibitem[{{Komatsu} \& {Seljak}(2002)}]{Komatsu_Seljak2002}
{Komatsu}, E., \& {Seljak}, U. 2002,
  \href{http://dx.doi.org/10.1046/j.1365-8711.2002.05889.x}{\mnras},
  \href{http://adsabs.harvard.edu/abs/2002MNRAS.336.1256K}{336},
  \href{http://adsabs.harvard.edu/abs/2002MNRAS.336.1256K}{1256}

\bibitem[{Kravtsov(1999)}]{Kravtsov1999}
Kravtsov, A.~V. 1999, PhD thesis, New Mexico State Univ.

\bibitem[{Kravtsov {et~al.}(2002)Kravtsov, Klypin, \& Hoffman}]{Kravtsov2002}
Kravtsov, A.~V., Klypin, A., \& Hoffman, Y. 2002,
  \href{http://dx.doi.org/10.1086/340046}{\apj},
  \href{http://stacks.iop.org/0004-637X/571/i=2/a=563}{571},
  \href{http://stacks.iop.org/0004-637X/571/i=2/a=563}{563}

\bibitem[{{Lagan{\'a}} {et~al.}(2010){Lagan{\'a}}, {de Souza}, \&
  {Keller}}]{Lagana2010}
{Lagan{\'a}}, T.~F., {de Souza}, R.~S., \& {Keller}, G.~R. 2010,
  \href{http://dx.doi.org/10.1051/0004-6361/200911855}{\aap},
  \href{http://adsabs.harvard.edu/abs/2010A%26A...510A..76L}{510},
  \href{http://adsabs.harvard.edu/abs/2010A%26A...510A..76L}{A76}

\bibitem[{Lau {et~al.}(2009)Lau, Kravtsov, \& Nagai}]{Lau2009}
Lau, E.~T., Kravtsov, A.~V., \& Nagai, D. 2009,
  \href{http://dx.doi.org/10.1088/0004-637X/705/2/1129}{\apj},
  \href{http://stacks.iop.org/0004-637X/705/i=2/a=1129?key=crossref.7b2d49cb5d8c13775308e4c1861163e0}{705},
  \href{http://stacks.iop.org/0004-637X/705/i=2/a=1129?key=crossref.7b2d49cb5d8c13775308e4c1861163e0}{1129}

\bibitem[{Lau {et~al.}(2013)Lau, Nagai, \& Nelson}]{Lau2013}
Lau, E.~T., Nagai, D., \& Nelson, K. 2013,
  \href{http://dx.doi.org/10.1088/0004-637X/777/2/151}{\apj},
  \href{http://arxiv.org/abs/1306.3993
  http://stacks.iop.org/0004-637X/777/i=2/a=151?key=crossref.2a43b61df90946fe754e21033777fe24}{777},
  \href{http://arxiv.org/abs/1306.3993
  http://stacks.iop.org/0004-637X/777/i=2/a=151?key=crossref.2a43b61df90946fe754e21033777fe24}{151}

\bibitem[{Mahdavi {et~al.}(2013)Mahdavi, Hoekstra, Babul, Bildfell, Jeltema, \&
  Henry}]{Mahdavi2013}
Mahdavi, A., Hoekstra, H., Babul, A., {et~al.} 2013,
  \href{http://dx.doi.org/10.1088/0004-637X/767/2/116}{\apj},
  \href{http://stacks.iop.org/0004-637X/767/i=2/a=116?key=crossref.52cc0b20a07f6ab3d1676119be697cec}{767},
  \href{http://stacks.iop.org/0004-637X/767/i=2/a=116?key=crossref.52cc0b20a07f6ab3d1676119be697cec}{116}

\bibitem[{Nagai {et~al.}(2003)Nagai, Kravtsov, \& Kosowsky}]{Nagai2003}
Nagai, D., Kravtsov, A.~V., \& Kosowsky, A. 2003,
  \href{http://dx.doi.org/10.1086/368281}{\apj},
  \href{http://stacks.iop.org/0004-637X/587/i=2/a=524}{587},
  \href{http://stacks.iop.org/0004-637X/587/i=2/a=524}{524}

\bibitem[{Nagai {et~al.}(2007{\natexlab{a}})Nagai, Kravtsov, \&
  Vikhlinin}]{Nagai2007a}
Nagai, D., Kravtsov, A.~V., \& Vikhlinin, A. 2007{\natexlab{a}},
  \href{http://dx.doi.org/10.1086/521328}{\apj},
  \href{http://stacks.iop.org/0004-637X/668/i=1/a=1}{668},
  \href{http://stacks.iop.org/0004-637X/668/i=1/a=1}{1}

\bibitem[{Nagai {et~al.}(2007{\natexlab{b}})Nagai, Vikhlinin, \&
  Kravtsov}]{Nagai2007b}
Nagai, D., Vikhlinin, A., \& Kravtsov, A.~V. 2007{\natexlab{b}},
  \href{http://dx.doi.org/10.1086/509868}{\apj},
  \href{http://stacks.iop.org/0004-637X/655/i=1/a=98}{655},
  \href{http://stacks.iop.org/0004-637X/655/i=1/a=98}{98}

\bibitem[{Nelson {et~al.}(2014)Nelson, Lau, Nagai, Rudd, \& Yu}]{Nelson2013}
Nelson, K., Lau, E.~T., Nagai, D., Rudd, D.~H., \& Yu, L. 2014,
  \href{http://dx.doi.org/10.1088/0004-637X/782/2/107}{\apj},
  \href{http://iopscience.iop.org/0004-637X/782/2/107/}{782},
  \href{http://iopscience.iop.org/0004-637X/782/2/107/}{107}

\bibitem[{Nelson {et~al.}(2012)Nelson, Rudd, Shaw, \& Nagai}]{Nelson2012}
Nelson, K., Rudd, D.~H., Shaw, L., \& Nagai, D. 2012,
  \href{http://dx.doi.org/10.1088/0004-637X/751/2/121}{\apj},
  \href{http://stacks.iop.org/0004-637X/751/i=2/a=121?key=crossref.7e56fe90276ff11ccb925b2ccbfaeb63}{751},
  \href{http://stacks.iop.org/0004-637X/751/i=2/a=121?key=crossref.7e56fe90276ff11ccb925b2ccbfaeb63}{121}

\bibitem[{Parrish {et~al.}(2012)Parrish, McCourt, Quataert, \&
  Sharma}]{Parrish2012}
Parrish, I.~J., McCourt, M., Quataert, E., \& Sharma, P. 2012,
  \href{http://dx.doi.org/10.1111/j.1745-3933.2011.01171.x}{\mnras},
  \href{http://arxiv.org/abs/1109.1285
  http://mnrasl.oxfordjournals.org/cgi/doi/10.1111/j.1745-3933.2011.01171.x}{419},
  \href{http://arxiv.org/abs/1109.1285
  http://mnrasl.oxfordjournals.org/cgi/doi/10.1111/j.1745-3933.2011.01171.x}{L29}

\bibitem[{Piffaretti \& Valdarnini(2008)}]{Piffaretti2008}
Piffaretti, R., \& Valdarnini, R. 2008,
  \href{http://dx.doi.org/10.1051/0004-6361:200809739}{\aap},
  \href{http://www.aanda.org/10.1051/0004-6361:200809739}{491},
  \href{http://www.aanda.org/10.1051/0004-6361:200809739}{71}

\bibitem[{Rasia {et~al.}(2004)Rasia, Tormen, \& Moscardini}]{Rasia2004}
Rasia, E., Tormen, G., \& Moscardini, L. 2004,
  \href{http://dx.doi.org/10.1111/j.1365-2966.2004.07775.x}{\mnras},
  \href{http://mnras.oxfordjournals.org/cgi/doi/10.1111/j.1365-2966.2004.07775.x}{351},
  \href{http://mnras.oxfordjournals.org/cgi/doi/10.1111/j.1365-2966.2004.07775.x}{237}

\bibitem[{Rasia {et~al.}(2006)Rasia, Ettori, Moscardini, Mazzotta, Borgani,
  Dolag, Tormen, Cheng, \& Diaferio}]{Rasia2006}
Rasia, E., Ettori, S., Moscardini, L., {et~al.} 2006,
  \href{http://dx.doi.org/10.1111/j.1365-2966.2006.10466.x}{\mnras},
  \href{http://mnras.oxfordjournals.org/cgi/doi/10.1111/j.1365-2966.2006.10466.x}{369},
  \href{http://mnras.oxfordjournals.org/cgi/doi/10.1111/j.1365-2966.2006.10466.x}{2013}

\bibitem[{Rudd {et~al.}(2008)Rudd, Zentner, \& Kravtsov}]{Rudd2008}
Rudd, D.~H., Zentner, A.~R., \& Kravtsov, A.~V. 2008,
  \href{http://dx.doi.org/10.1086/523836}{\apj},
  \href{http://stacks.iop.org/0004-637X/672/i=1/a=19}{672},
  \href{http://stacks.iop.org/0004-637X/672/i=1/a=19}{19}

\bibitem[{Shaw {et~al.}(2010)Shaw, Nagai, Bhattacharya, \& Lau}]{Shaw2010}
Shaw, L.~D., Nagai, D., Bhattacharya, S., \& Lau, E.~T. 2010,
  \href{http://dx.doi.org/10.1088/0004-637X/725/2/1452}{\apj},
  \href{http://stacks.iop.org/0004-637X/725/i=2/a=1452?key=crossref.9e3ce4a68e9e9de20cb7752faa42acf8}{725},
  \href{http://stacks.iop.org/0004-637X/725/i=2/a=1452?key=crossref.9e3ce4a68e9e9de20cb7752faa42acf8}{1452}

\bibitem[{{Shi} \& {Komatsu}(2014)}]{Shi2014}
{Shi}, X., \& {Komatsu}, E. 2014, \mnras, submitted,
  arXiv:\href{http://adsabs.harvard.edu/abs/2014arXiv1401.7657S}{1401.7657}

\bibitem[{{Suto} {et~al.}(2013){Suto}, {Kawahara}, {Kitayama}, {Sasaki},
  {Suto}, \& {Cen}}]{Suto2013}
{Suto}, D., {Kawahara}, H., {Kitayama}, T., {et~al.} 2013,
  \href{http://dx.doi.org/10.1088/0004-637X/767/1/79}{\apj},
  \href{http://adsabs.harvard.edu/abs/2013ApJ...767...79S}{767},
  \href{http://adsabs.harvard.edu/abs/2013ApJ...767...79S}{79}

\bibitem[{Takahashi {et~al.}(2010)Takahashi, Mitsuda, Kelley, Aharonian,
  Akimoto, Allen, Anabuki, Angelini, Arnaud, Awaki, Bamba, Bando, Bautz,
  Blandford, Boyce, Brown, Chernyakova, Coppi, Costantini, Cottam, Crow,
  de~Plaa, de~Vries, den Herder, DiPirro, Done, Dotani, Ebisawa, Enoto, Ezoe,
  Fabian, Fujimoto, Fukazawa, Funk, Furuzawa, Galeazzi, Gandhi, Gendreau,
  Gilmore, Haba, Hamaguchi, Hatsukade, Hayashida, Hiraga, Hirose,
  Hornschemeier, Hughes, Hwang, Iizuka, Ishibashi, Ishida, Ishimura, Ishisaki,
  Isobe, Ito, Iwata, Kaastra, Kallman, Kamae, Katagiri, Kataoka, Katsuda,
  Kawaharada, Kawai, Kawasaki, Khangaluyan, Kilbourne, Kinugasa, Kitamoto,
  Kitayama, Kohmura, Kokubun, Kosaka, Kotani, Koyama, Kubota, Kunieda, Laurent,
  Lebrun, Limousin, Loewenstein, Long, Madejski, Maeda, Makishima, Markevitch,
  Matsumoto, Matsushita, McCammon, Miller, Mineshige, Minesugi, Miyazawa,
  Mizuno, Mori, Mori, Mukai, Murakami, Murakami, Mushotzky, Nakagawa, Nakagawa,
  Nakajima, Nakamori, Nakazawa, Namba, Nomachi, O'Dell, Ogawa, Ogawa, Ogi,
  Ohashi, Ohno, Ohta, Okajima, Ota, Ozaki, Paerels, Paltani, Parmar, Petre,
  Pohl, Porter, Ramsey, Reynolds, Sakai, Sambruna, Sato, Sato, Serlemitsos,
  Shida, Shimada, Shinozaki, Shirron, Smith, Sneiderman, Soong, Stawarz,
  Sugita, Szymkowiak, Tajima, Takahashi, Takei, Tamagawa, Tamura, Tamura,
  Tanaka, Tanaka, Tanaka, Tashiro, Tawara, Terada, Terashima, Tombesi, Tomida,
  Tozuka, Tsuboi, Tsujimoto, Tsunemi, Tsuru, Uchida, Uchiyama, Uchiyama, Ueda,
  Uno, Urry, Watanabe, White, Yamada, Yamaguchi, Yamaoka, Yamasaki, Yamauchi,
  Yamauchi, Yatsu, Yonetoku, \& Yoshida}]{Takahashi2010}
Takahashi, T., Mitsuda, K., Kelley, R., {et~al.} 2010, in Society of
  Photo-Optical Instrumentation Engineers (SPIE) Conference Series, ed.
  M.~Arnaud, S.~S. Murray, \& T.~Takahashi, Vol. 7732, 77320Z,
  arXiv:\href{http://proceedings.spiedigitallibrary.org/proceeding.aspx?articleid=1346337}{1010.4972}

\bibitem[{Tinker {et~al.}(2008)Tinker, Kravtsov, Klypin, Abazajian, Warren,
  Yepes, Gottl\"{o}ber, \& Holz}]{Tinker2008}
Tinker, J., Kravtsov, A.~V., Klypin, A., {et~al.} 2008,
  \href{http://dx.doi.org/10.1086/591439}{\apj},
  \href{http://stacks.iop.org/0004-637X/688/i=2/a=709}{688},
  \href{http://stacks.iop.org/0004-637X/688/i=2/a=709}{709}

\bibitem[{{Trac} {et~al.}(2011){Trac}, {Bode}, \& {Ostriker}}]{Trac2011b}
{Trac}, H., {Bode}, P., \& {Ostriker}, J.~P. 2011,
  \href{http://dx.doi.org/10.1088/0004-637X/727/2/94}{\apj},
  \href{http://adsabs.harvard.edu/abs/2011ApJ...727...94T}{727},
  \href{http://adsabs.harvard.edu/abs/2011ApJ...727...94T}{94}

\bibitem[{Vazza {et~al.}(2011)Vazza, Brunetti, Gheller, Brunino, \&
  Br\"{u}ggen}]{Vazza2011}
Vazza, F., Brunetti, G., Gheller, C., Brunino, R., \& Br\"{u}ggen, M. 2011,
  \href{http://dx.doi.org/10.1051/0004-6361/201016015}{\aap},
  \href{http://www.aanda.org/10.1051/0004-6361/201016015}{529},
  \href{http://www.aanda.org/10.1051/0004-6361/201016015}{A17}

\bibitem[{{von der Linden} {et~al.}(2014){von der Linden}, {Mantz}, {Allen},
  {Applegate}, {Kelly}, {Morris}, {Wright}, {Allen}, {Burchat}, {Burke},
  {Donovan}, \& {Ebeling}}]{vonderLinden2014}
{von der Linden}, A., {Mantz}, A., {Allen}, S.~W., {et~al.} 2014, \mnras,
  submitted,
  arXiv:\href{http://adsabs.harvard.edu/abs/2014arXiv1402.2670V}{1402.2670}

\bibitem[{{Zandanel} \& {Ando}(2014)}]{Zandanel2014}
{Zandanel}, F., \& {Ando}, S. 2014,
  \href{http://dx.doi.org/10.1093/mnras/stu324}{\mnras},
  \href{http://adsabs.harvard.edu/abs/2014MNRAS.440..663Z}{440},
  \href{http://adsabs.harvard.edu/abs/2014MNRAS.440..663Z}{663}

\bibitem[{Zhang {et~al.}(2010)Zhang, Okabe, Finoguenov, Smith, Piffaretti,
  Valdarnini, Babul, Evrard, Mazzotta, Sanderson, \& Marrone}]{Zhang2010}
Zhang, Y.-Y., Okabe, N., Finoguenov, A., {et~al.} 2010,
  \href{http://dx.doi.org/10.1088/0004-637X/711/2/1033}{\apj},
  \href{http://stacks.iop.org/0004-637X/711/i=2/a=1033?key=crossref.4ef862369b1da85be8a8160a1bf933ef}{711},
  \href{http://stacks.iop.org/0004-637X/711/i=2/a=1033?key=crossref.4ef862369b1da85be8a8160a1bf933ef}{1033}

\bibitem[{{Zhuravleva} {et~al.}(2013){Zhuravleva}, {Churazov}, {Kravtsov},
  {Lau}, {Nagai}, \& {Sunyaev}}]{Zhuravleva2013}
{Zhuravleva}, I., {Churazov}, E., {Kravtsov}, A., {et~al.} 2013,
  \href{http://dx.doi.org/10.1093/mnras/sts275}{\mnras},
  \href{http://adsabs.harvard.edu/abs/2013MNRAS.428.3274Z}{428},
  \href{http://adsabs.harvard.edu/abs/2013MNRAS.428.3274Z}{3274}

\bibitem[{Zhuravleva {et~al.}(2011)Zhuravleva, Churazov, Sazonov, Sunyaev, \&
  Dolag}]{Zhuravleva2011}
Zhuravleva, I.~V., Churazov, E.~M., Sazonov, S.~Y., Sunyaev, R.~A., \& Dolag,
  K. 2011, \href{http://dx.doi.org/10.1134/S1063773711010087}{Astron. Lett.},
  \href{http://arxiv.org/abs/1102.4098
  http://link.springer.com/10.1134/S1063773711010087}{37},
  \href{http://arxiv.org/abs/1102.4098
  http://link.springer.com/10.1134/S1063773711010087}{141}

\end{thebibliography}

\appendix
\label{sec:appendix}

\section{A. Fitting the Non-thermal Pressure Fraction Profile with Respect to Critical Density}
\label{sec:scaling}

\begin{figure*}[t]
\begin{center}  
\includegraphics[scale=0.85]{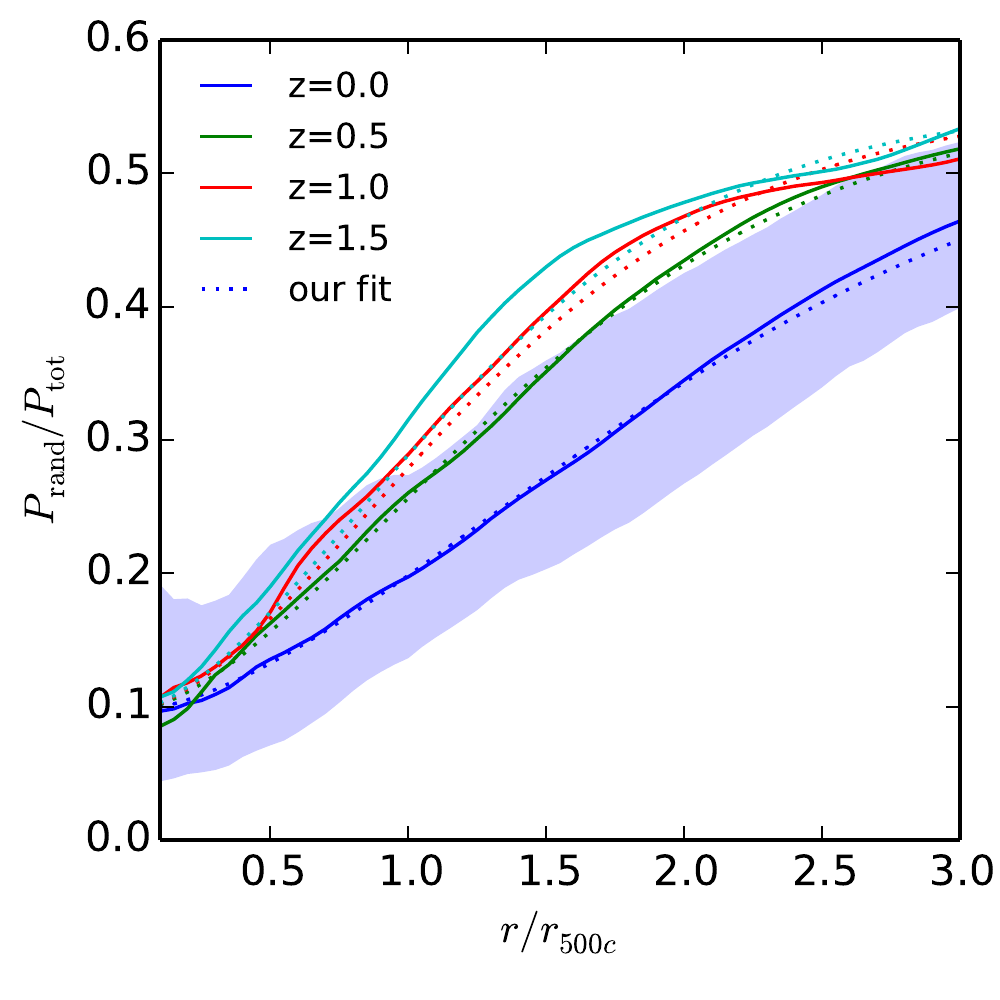}
\includegraphics[scale=0.85]{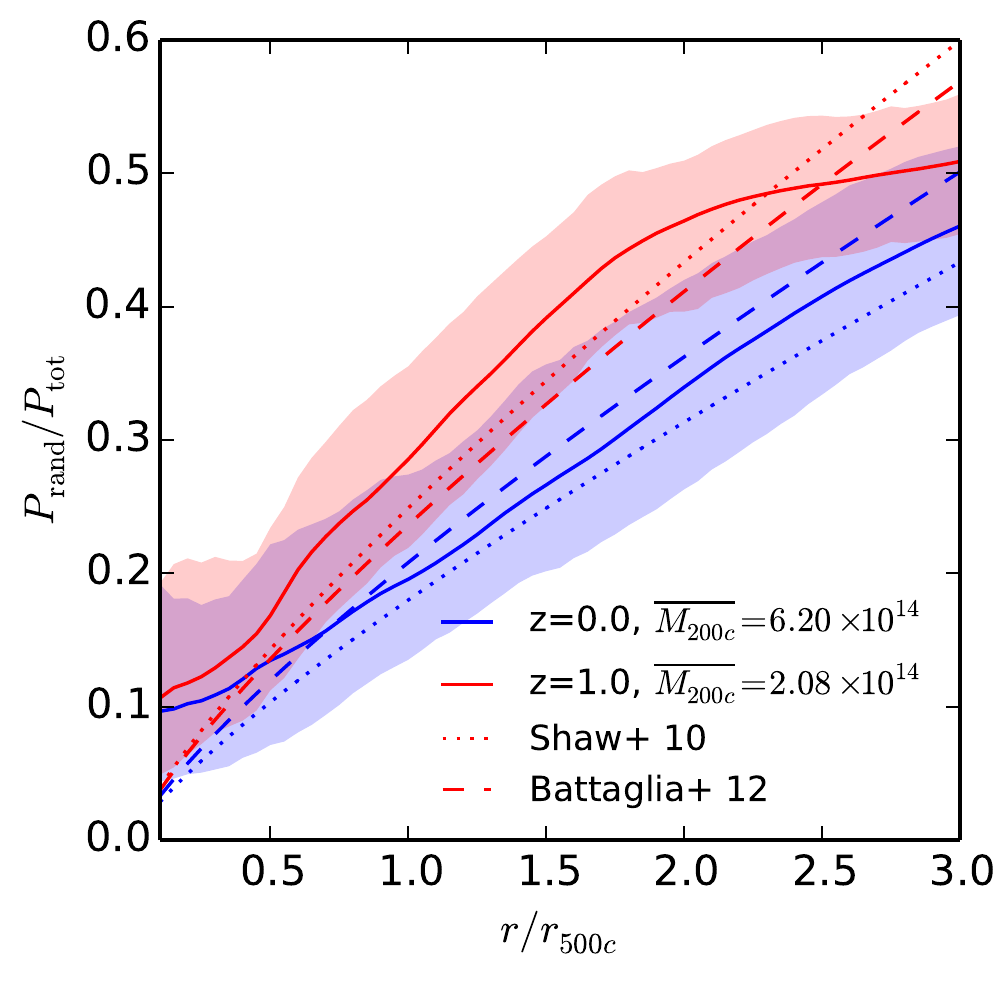}
\caption
{Left: $\pfrac$ profiles for scale radii with respect to critical density, $r_{500c}$.  The figure shows the redshift dependence across four redshifts, $z=0.0,0.5,1.0, 1.5$ in blue, green, red and cyan lines, respectively. The mean profiles for our sample are shown in the solid lines and the shaded regions denote the 1-$\sigma$ scatter around the mean at $z=0$. The scatter for the other redshifts is comparable and has been omitted for clarity. The dotted lines depict our converted fitting formula. Right: Comparison of the \cite{Shaw2010} (dotted) and \cite{Battaglia2012a} (dashed) fitting formulas with our data at $z=0$ and $z=1$, blue and red lines respectively. The shaded regions denote the 1-$\sigma$ scatter around mean at each redshift.}
\label{fig:p_otherradii}
\end{center}
\end{figure*}

To date, it has been common to use the critical density to define the radius and mass of galaxy clusters. We therefore provide a conversion for our fitting formula for the universal non-thermal pressure profile defined respect to the mean mass density of the universe, so that it can easily be used with alternative definitions of halos' radius and mass based on the critical density. This conversion method is adapted from Appendix C in \cite{Hu2003}. The converted fitting formula is expressed as 
\begin{equation}
\frac{P_{\rm rand}}{P_{\rm total}} = 1-A\ \left\{1+\exp\left[-\left(\frac{\eta r/r_{200m}}{B}\right)^{\gamma}\right]\right\}
\end{equation}
where $A$, $B$, and $\gamma$ are the best fit parameters given in Section~\ref{function}. The conversion factor $\eta$ is given by 
\begin{equation}
\eta = c_{200m} \left( \frac{1}{ \sqrt{   a_0f^{2p} +(3/4)^2  }}+ 2f \right) 
\end{equation}
where $c_{200m} \equiv r_{200m}/r_s$ is the concentration parameter, and
\begin{align}      
f &= \frac{\ln(1+c_{200m}) - c_{200m}/(1+c_{200m})}{c_{200m}^3} \left(\frac{\Delta_c}{200} \frac{E(z)^2}{\Omega_m(1+z)^3} \right), \\
p &= a_1+a_2\ln f + a_3\left(\ln f\right)^2.  
\end{align}
Here $\Delta_c$ is the chosen overdensity with respect to the critical density of the universe, $E(z)\equiv H(z)/H(z=0) =  \sqrt{\Omega_m(1+z)^3+\Omega_{\Lambda}}$ is the normalized Hubble parameter for a flat cosmology, and $a_0 = 0.5116$, $a_1 = -1.285/3$, $a_2 = -3.13\times10^{-3}$ and $a_3 = -3.52\times 10^{-5}$. 

In left panel of Figure~\ref{fig:p_otherradii}, we show how our fitting formula recovers the non-thermal pressure fraction measured from simulations at four different redshifts, $z= 0.0,0.5,1.0,1.5$.  To convert between $r_{200m}$ and $r_{500c}$ we use the mean concentration at each redshift bin as calculated from concentration-mass relation in \citet{Bhattacharya2011}. We find excellent agreement between our fit and our data in the redshift range of $0\leq z \leq 1$. At $z=1.5$ our conversion underestimates the true profile out to 2$r_{500c}$. 

Previous attempts to describe the non-thermal pressure profile used the scale radius $r_{500c}$. In right panel of Figure~\ref{fig:p_otherradii} we compare our sample to the fitting formula from \cite{Battaglia2012a} (dashed line) and \cite{Shaw2010} (dotted line) at two redshifts, $z=0.0$ and $z=1.0$, shown in the blue and red lines, respectively.  The \cite{Shaw2010} fitting formula slightly underestimates our results at $z=0$. At $z=1.0$, the fit has approximately the same slope as ours between $r=0.5 r_{500c}$ and $r=1.5 r_{500c}$, but underestimates the pressure fraction slightly at $r\lesssim 2.2 r_{500c}$ and overestimates the fraction at the larger radii. The deviation in \cite{Shaw2010} is likely due to the fact that their cluster sample was small (only 16) and heterogeneous. The \cite{Battaglia2012a} fitting formula adopts the redshift dependence from \cite{Shaw2010} and adds an additional factor to account for the mass dependence which they calibrate using a larger sample of simulated clusters at $z=0$. At $z=0$, this additional factor results in a slight overestimate of the non-thermal pressure fraction out to $r=2r_{500c}$.  At $z=1$, their fit shows the same redshift dependence as \cite{Shaw2010}, but with a lower normalization due to the mass dependence factor, which was only calibrated at $z=0$.

\begin{figure*}
\begin{center}  
\includegraphics[scale=0.7]{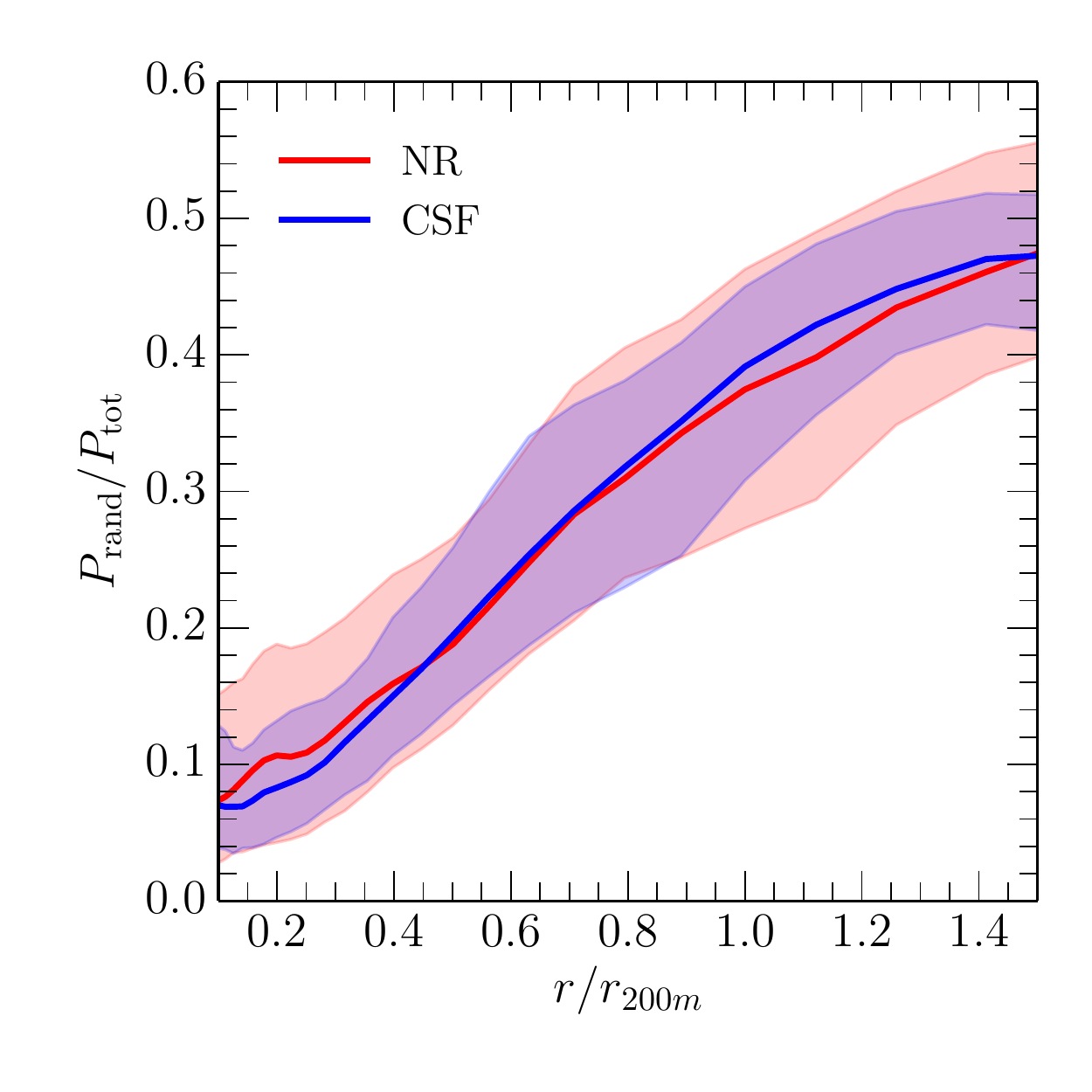}
\includegraphics[scale=0.7]{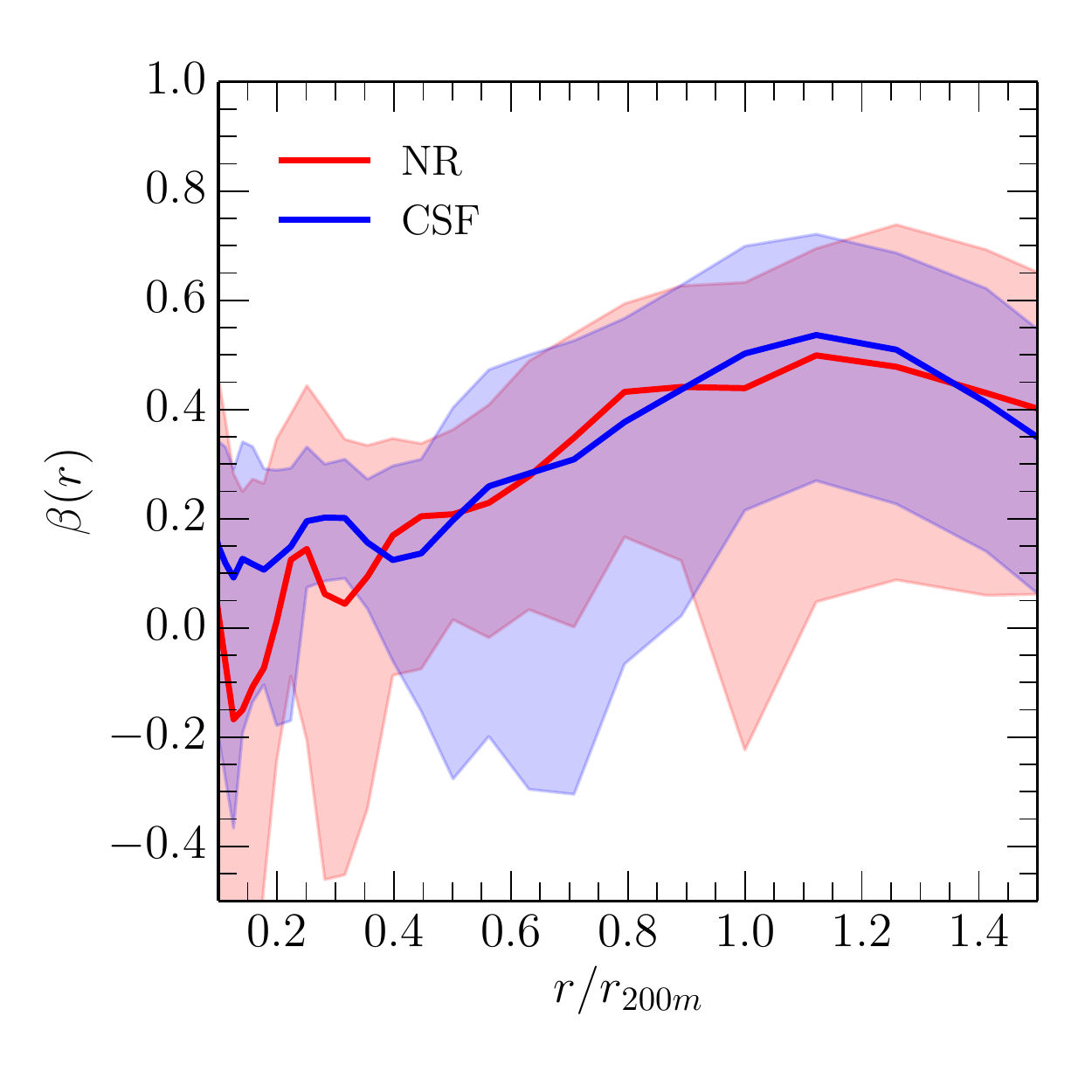}
\caption
{Left: $\pfrac$ profiles for the 16 clusters taken from \citet{Nagai2007a}.
Right: The gas velocity anisotropy profiles $\beta$ for the same set of clusters. 
The red lines correspond to clusters simulated with non-radiative (NR) physics, and blue lines correspond to the same clusters simulated with radiative cooling, star formation and supernova feedback (CSF). The shaded regions indicate 1-$\sigma$ scatter around the mean. }
\label{fig:physics}
\end{center}
\end{figure*}

\section{B. Effects of Dissipative Physics on the Non-thermal Pressure Fraction and Gas Velocity Anisotropy}
\label{sec:physics}

In this section we examine the effect of dissipative physics on the non-thermal pressure fraction and gas velocity anisotropy in group and cluster sized systems. We compare profiles for the set of 16 clusters from \citet{Nagai2007a}, simulated with two different gas physics: non-radiative (NR) physics; and with radiative cooling, star formation, and supernova feedback (CSF). In Figure~\ref{fig:physics} we show the mean profiles of the non-thermal pressure fraction  $\pfrac$ and the gas velocity anisotropy $\beta$ for the clusters at $z=0$. There is no systematic dependence on gas physics in the radial range of $0.1 \lesssim r/r_{200m} \lesssim 1.5$.  The profiles between the two runs are consistent between each other within 1-$\sigma$. The addition of dissipative physics has little effect on both the non-thermal pressure fraction and the gas velocity anisotropy in this radial range.

\end{document}